\documentstyle[12pt]{article}
\newcommand{\be}{\begin{equation}}
\newcommand{\ee}{\end{equation}}
\newcommand{\ba}{\begin{array}{l}}
\newcommand{\ea}{\end{array}}
\newcommand{\bqa}{\begin{eqnarray}}
\newcommand{\eqa}{\end{eqnarray}}
\newcommand{\bb}{\begin{thebibliography}}
\newcommand{\eb}{\end{thebibliography}}
\newcommand{\Tr }{\mbox{Tr}\;}

\begin{document}

\begin{titlepage}
\begin{center}
{\bf   RESEARCH INSTITUTE OF APPLIED PHYSICS \\
       TASHKENT STATE UNIVERSITY} \\
\vspace{20mm}
\hfill{\bf Preprint NIIPF-92/01}
\vspace{60mm}

{\bf \Large MINIMAL DEFORMATIONS OF THE COMMUTATIVE \\
 ALGEBRA AND THE LINEAR GROUP GL(n) }\\
\vspace{20mm}
 {\bf B.M.ZUPNIK} \\
\vspace{70mm}
{\bf Tashkent - 1992}
\end{center}
\end{titlepage}

\newpage
\begin{titlepage}
\hfill Preprint NIIPF-92/01
\hfill April 1992
\vspace{40mm}
\begin{center}
{\bf \Large MINIMAL DEFORMATIONS OF THE COMMUTATIVE \\
 ALGEBRA AND THE LINEAR GROUP GL(n) }\\
\vspace{20mm}
 {\bf B.M.ZUPNIK} \\
\vspace{10mm}
{\bf   Research Institute of Applied Physics, \\
       Tashkent State University, Vuzgorodok, \\
   Tashkent 700095, Uzbekistan} \\
\end{center}
\vspace{20mm}

\begin{abstract}
We consider the relations of generalized commutativity in the algebra of
formal series $ M_q (x^i ) $, which conserve a tensor $ I_q $-grading and
depend on parameters $ q(i,k)  $ . We choose the $ I_q $-preserving version of
differential calculus on $ M_q$ . A new construction of the symmetrized tensor
product for $ M_q $-type algebras and the corresponding definition of minimally
deformed linear group $ QGL(n) $ and Lie algebra $ qgl(n) $ are proposed. We
study the connection of $ QGL(n) $ and $ qgl(n) $ with the special matrix
algebra $ \mbox{Mat} (n,Q) $ containing matrices with noncommutative elements.
A definition of the deformed determinant in the algebra $ \mbox{Mat} (n,Q) $ is
given. The exponential parametrization in the algebra $ \mbox{Mat} (n,Q) $  is
considered on the basis of Campbell-Hausdorf formula.  \end{abstract}
\end{titlepage} \newpage

\section{Introduction.The principle of minimal deformation} \indent

Mathematical formalism of the quantum inverse scattering problem can be
constructed on the basis of a concept of quantum group, which is developed
intensively as a new branch of modern mathematics [1-4]. Quantum groups are
considered as parametric deformations of the classical groups.  Representations
of quantum groups are closely connected with deformations of commutative
algebras. Let $ x^1 ,\ldots x^n $ are generators of the formal-series algebra
over a field of comlex numbers $ {\bf C}  $ . A deformation of the commutative
algebra can be defined with the help of the following bilinear relations \be
x^i\; x^j =\hat{R}^{ij}_{kl}\;x^k\; x^l \label{a1} \ee where $ \hat{R}\;
\mbox{is}\; n^2\times n^2 \;$ matrix which satisfies the constant Yang-Baxter
equation:  \be \hat{R}^{ij}_{rt}\;\hat{R}^{tn}_{pl}\;
\hat{R}^{rp}_{mk}\;=\;\hat{R}^{ip}_{mr}\; \hat{R}^{jn}_{pt}
\;\hat{R}^{rt}_{kl}\label{a2} \ee

These relations can be written in terms of the $ R $-matrix \be
R^{ij}_{kl}\;=\;(P\hat{R})^{ij}_{kl}=\hat{R}^{ji}_{kl}\label{a3} \ee

We shall consider a limited class of deformations corresponding to the diagonal
$ R $-matrices \be R^{ji}_{kl}\;=\;q(i,j)\;\delta_k^j \; \delta^i_l
=\hat{R}^{ij}_{kl}\label{a4} \ee where $ q(i,j) \in{ \bf C}  $ are the
parameters of deformations.

Let us define the $ q $-deformed algebra $ M_q (n) $ (quantum space or $ q
$-commutative algebra ) as the formal-series algebra with generatots $ x^i $
satisfying the following relations \bqa & x^i\; x^j \;=\; q(i,j)\;x^j \;x^i
\;=\;[ij]\;x^j \;x^i & \label{a5} \\ & q(i,i)=1,\;\;\;\; q(i,j)\;q(j,i)=1 &
\label{a6} \eqa

Here the notation $ [ij]=q(i,j) $ is introduced. The algebra  $ M_q (n) $ has $
(n^2 - n)/2 $ independent parameters.

Note, that we shall use a summation convention for coinciding upper and low
indices only. We do not treat symbols in parentheses as indeces, so summation
is not used in Eqs (\ref{a5} ,\ref{a6}).

The standard ${\bf Z} $-grading in the algebra of formal series corresponds to
a decomposition in degrees $ p $ of the monomials \be x^{i_1}\;x^{i_2}\cdots
x^{i_p}\;\stackrel{\rm def}{=}\;x^{I(p)}\label{a7} \ee

Consider a set $ I $ of the totally symmetric multiindeces \be I(p)=(i_1 ,i_2
\ldots  i_p)\label{a8} \ee

It is easy to introduce a commutative multiplication of multiindeces in $ I $
\be I(p)\ast I(r)=I(p+r)=I(r)\ast I(p)\label{a9} \ee

A unit element $ I(0) $ (zero multiindex) of the commutative semigroup $ I $
corresponds to absence of indices.

Consider a generalized tensor $I_q$-grading as decomposition of the algebra
$M_q (n)$ in a direct sum of vector spaces $ M_q (I(p)) $ corresponding to
monomials (\ref{a7}) with an arbitrary order of indeces \[ M_q (n)=
 \bigoplus_{p=0}^{\infty}\;\bigoplus_{I(p)}\;  M_q (I(p))=\;
\bigoplus_{p=0}^{\infty}\;\bigoplus_{I(p)}\; \left\{ kx^{I(p)}\right\} \] \be
 M_q (I(p))\;M_q (I(r))\;\subset\; M_q (I(p+r))  \label{a10} \ee

$I_q$-grading in the $ q $-commutative algebra $ M_q (n) $ is consistent with a
simple generalization of the relation (\ref{a5}) to the arbitrary monomials \be
 x^{I(p)}\;x^{K(r)}=q(I,K)\;x^{K(r)}\;x^{I(p)} \label{a11} \ee \be
  q(I,K)\;\stackrel{\rm def}{=}\;[I(p)|K(r)]=\; \prod_{\alpha=1}^{p}\;
 \prod_{\beta=1}^{r}\; \left[ i_\alpha \;k_\beta \right]   \label{a12} \ee \bqa
 & q(I,K)\;q(K,I)=1,\;\;\;q(I,I)=1 &\nonumber \\ & q(I\ast
 J,\;K)=q(I,K)\;q(J,K) & \label{a13} \eqa

One can formulate mnemonic rule for the calculation of the function $ q(I,K) $:
{\bf The multiplier $ q(i,k) $ appears when the index $ i $ moves by the index
$ k $ from the left to right}. This rule is analogous to the rule of signs in
${\bf Z_2} $ -graded algebras [5,6], which is a key principle in the
supersymmetric generalization of the "commutative" algebra and analysis. In
further considerations this rule can be generalized taking into account the
 introducing of  covariant (low) indeces and other extensions of $ I_q
 $-grading.

In Ref[7] and other works there were considered $ (G,f) $-graded algebras which
correspond to Abelian groups $ G $ and commutation functions $ f(g,h) $
satisfying the  Eq(\ref{a13})-type restrictions. We shall not use the accepted
in mathematical works name "coloured" for these qroups and algebras, because
this word is widely used in the quantum chromodynamics.

According to the results of Ref[7] $ (G,f) $-Lie algebras for the finite groups
$ G $ can be reduced to ordinary Lie algebras or Lie superalgebras if the
restrictions $ f(g,g)=\pm 1 $ is used.

Note, that  deformations of formal Lie groups and Lie algebras corresponding to
a general $ \hat{R} $-matrix and condition $ \hat{R}^2 =1 $ was investigated in
Ref[8].

We shall treat the algebra $ M_q (n) $ as {\bf minimal deformation } of the
commutative formal-series algebra $  C (x^i) $ and shall use the principle of
minimal deformation for constructing the theories, which consistent with $ I_q
$-grading. One can see from Refs[7-11] that there exists some uncertainty in
the constructing of a differential calculus and the action of a quantum linear
group on the algebra $ M_q (n) $ . It seems to us very natural to build these
theories on a basis of the minimal deformation  (MD)  principle.

One can made a linear similarity transformation with the complex matrix $ T_k^i
$ [14] in the algebra $ M_q (n) $ , which does not conserve $ I_q $ -grading.It
is evident that this transformation generate undiagonal solution of
 Eq(\ref{a2}), which is similar to the solution (\ref{a4}). Note that the
diagonal solution (\ref{a4}) is invariant under the transformation $ T_k^i =
t(i)\delta_k^i $ . An arbitrary transformation $ T_k^i $ conserves the
following relations for the matrix (\ref{a4}):  \bqa &
\hat{R}_{kl}^{ij}\;\hat{R}_{mn}^{kl}=\delta_m^i \;\delta_n^j,\;\;\;
\hat{R}_{kj}^{ki}=\delta_j^i=\hat{R}_{jk}^{ik} &\nonumber \\ &
\hat{R}_{lm}^{ij}\;\hat{R}_{ik}^{lp}\;=\;\delta_k^j \;\delta_m^p =
 \hat{R}_{ml}^{ji}\; \hat{R}_{ki}^{pl} & \label{a14} \eqa

A generalization of $ I_q $-grading and a differential calculus on the algebra
$ M_q (n) $ are considered in section 2. Section 3 contains a definition of the
minimal deformation $ QGL(n) $ for the linear group and its connection with the
deformed Lie algebra $ qgl(n) $ . An alternative definition of the quantum
group $ QGL(n) $  in terms of the special matrix algebra $ \mbox{Mat}(n,Q) $ is
suggested in section 5.

It should be stressed, that our main purpose is the constructive discussion of
the simple formalism of deformations, which may be convenient for physical
applications, so we do not give a detailed review of references on quantum
groups and construct proofs of basic statements on the level of strictness
accepted in theoretical physics. The standard definition of quantum groups is
based on the original formulation of the inverse scattering problem [1]. We do
not know how to use minimal deformations of the linear group in physics, but a
search of corresponding applications seems to us very interesting.

In the conclusion we discuss a solution of the quantum Yang-Baxter equation,
which depends on the functions $ q(i,k,u,h) $ , where $ u $ is a spectral
parameter and $ h $ is a quasi-classical parameter.

Note, that minimal deformations of the linear supergroup $ GL(p,q) $ can be
constructed by the analogy with $ QGL(n) $.

\setcounter{equation}{0} \section{Differential calculus on minimally deformed
formal-series algebras} \indent

Consider the algebra $ M^*_q (n) $ with generations $ y_i $ satisfying the
relation \be y_i\; y_j =[ ij ]\; y_j\; y_i \label{b1} \ee

The $ I_q $-grading of monomials $ y_{k_1} \ldots  y_{k_p} =y_{K(p)} $ is
determined by covariant (low) multiindex $ K(p) $. A multiplication in the
algebra $ M_q^* (n) $ is consistent with the principle of minimal deformation
\be y_{K(p)}\; y_{K(r)}=\left[ K(p)|K(r)\right]\; y_{K(r)}\;y_{K(p)} \label{b2}
\ee where the commutation function can be determined by Eq(\ref{a12}).

Denote the special tensor product of the algebras $ M_q\; \mbox{and}\; M^*_q $
by the symbol $ M_q \bigotimes_q M^*_q =M_q (x^i , y_k ) $ . Let us treat $ M_q
(x,y) $ as the formal-series algebra over $ {\bf C} $ with $ 2n $ generators $
x^i , y_k $ , so the symbol $ \bigotimes_q $ can be omitted in many cases. A
multiplication of generators $ x^i , y_k $ can be determined by Eqs(\ref{a5} ,
\ref{b1}) and the relation \be y_k \;x^i =\left[ ik \right] \;x^i\; y_k
\label{b3} \ee

Special tensor products of several algebras $ M_q \bigotimes_q \ldots
\bigotimes_q M_q \bigotimes_q M^*_q \bigotimes_q \ldots  \bigotimes_q M^*_q =
\linebreak  M(x_\alpha ^i , y_{k\beta}) $ are defined as the algebra of formal
series with several sets of generators $ x_1^i \ldots  x_p^i ,\;\;\;\linebreak
y_{k1} \ldots  y_{kr} $ and the commutation relations independent of the
additional indices $ \alpha ,\beta $ \bqa & x_\alpha ^i\; x_\beta ^i=\left[ ij
\right]\; x_\beta ^j\; x_\alpha ^i,\;\;\;\; y_{i\alpha}\;y_{j\beta}=\left[ ij
\right]\; y_{j\beta}\;y_{i\alpha} & \nonumber \\ & x_\alpha ^i\; y_{k\beta} =
\left[ ki \right]\;  y_{k\beta}\; x_\alpha^i & \label{b4} \eqa

One can analogously define the special tensor product of $ I_q $-graded moduli
over the algebra $ M_q $ .

A standard definition of tensor product $ \bigotimes $ is not consistent with
 MD-principle \bqa & (x^i \bigotimes x^k )(x^l \bigotimes x^j )=(x^i\; x^l
\bigotimes x^k \;x^j ) &\nonumber \\ & (x^i \bigotimes y_k)(x^l \bigotimes
y_j)=(x^i \;x^l \bigotimes y_k \;y_j) & \label{b5} \eqa

It should be stressed, that the covariant definition of a tensor product in
${\bf Z_2 }$-graded algebras is consistent with the sign rule [6].

The choice of the relations (\ref{b1} ,\ref{b3}) follows from the requirement
that product of corresponding generators with equal indices $ x^1 y_1 , x^2 y_2
\ldots   $ must commute with any generators $ x^i , y_k $ :  \bqa & \left[ x^i,
x^1 y_1 \right] =0=\left[ x^i, x^2 y_2 \right]=\ldots   &\nonumber \\ & \left[
y_k, x^1 y_1 \right] =0=\left[ y_k, x^3 y_3 \right]=\ldots   & \label{b6} \eqa

A generalized $ \widetilde{I}_q $-grading for the homogeneous element $ z_K^I =
x^{I(p)}\;y_{K(r)} \mbox{of}\; M_q (x,y) $ can be defined by a table containing
upper and low multiindeces \be I\left(  \frac{p}{r}\right)  =\left[
\frac{I(p)}{K(r)}\right] = \left[ \frac{i_1 \ldots  i_p} {k_1 \ldots k_r}
\right] \label{b7} \ee

A commutation relation for the elements $ z_K^I $ has the following form \bqa &
z_K^I\; z_L^J = \left[ \frac{I|J}{K|L} \right]\; z_L^J\; z_K^I & \label{b8} \\
& \left[ \frac{I|J}{K|L} \right] = \left[ I|J \right]\; \left[ J|K \right]\;
 \left[ K|L \right] \;\left[ L|I \right] & \label{b9} \eqa where the notation
of repeating products of $ [ij] $ multipliers Eq(\ref{a12}) is used.

A semigroup $ \widetilde{I}_q $ is the set of pairs $ I\left(
\frac{p}{r}\right) $ of the symmetrized multiindeces (\ref{b7}) and the
multiplication in $ \widetilde{I}_q $ can be defined with the help of Eq
(\ref{a9}) :  \be I\left(  \frac{p}{r}\right)  \ast I\left(  \frac{s}{t}\right)
=\left[ \frac{I(p)\ast I(s)} {K(r)\ast K(t)} \right] =I\left(
\frac{p+s}{r+t}\right)   \label{b10} \ee

One can introduce the equivalence relation in the algebra $ M_q (x,y) $ . Let
us speak that the elements $ z_K^I\; \mbox{and}\;  z_N^P\; $ are equivalent if
the following relations are fulfilled for any pair of multiideces $ J,L $ :
\be \left[ \frac{I|J}{K|L} \right]=\left[ \frac{P|J}{N|L} \right] \label{b11}
\ee

It is easy to show that in a general case the equivalent elements differ by
addition of one or several pairs of coinciding upper and low indeces, for
example \be z_j^i \sim z_{j1}^{i1}\sim z_{j13}^{i13}\sim \ldots \label{b12} \ee

In particular, all diagonal elements with coinciding upper and low indeces
belong to the centre $ C(x,y) $ of the algebra $ M_q (x,y) $. Note, that the
equivalence of elements in $ M_q (x,y) $ does not mean the equivalence of their
cotransformation laws in the quantum group. A connection of different tensor
representation can be realized with the help of contraction operation.

One can see from Refs [9-11] that an introduction of partial derivatives in the
algebra $ M_q (n) $ is  uncertain procedure, because there is no accepted
generalization of the Leibnitz rule for differentiation on the algebra of
functions. Using of the MD-principle removes this uncertainty \bqa &
\partial_i\; x^k =\delta_i^k + \left[ ki \right]\; x^k \;\partial_i &\nonumber
\\ & \partial_i \;\partial_k =\left[ ik \right]\; \partial_k\; \partial_i &
\label{b13} \eqa

The $ I_q $-deformed external algebra $ \Lambda \left( M_q (n) \right) =
\Lambda_q (n) $ can  be defined as an algebra with generators $ x^i ,\xi^i=
dx^i $ , which satisfy the following relations \bqa & \xi^i\; \xi^k = - \left[
ik \right]\; \xi^k \;\xi^i &\nonumber \\ & x^i\; \xi^k =\left[ ik \right]\;
\xi^k\; x^i & \label{b14} \eqa

The algebra $ \Lambda_q (n) $ can be treated as a  modulus over $ M_q (n) $
which has an additional ${\bf Z_n } $-grading correspondingly to degrees of $
\xi^i $ . An operator of external derivation $ d=dx^i \partial_i $ [4,9] is
defined by relations \be d^2=0,\;\;\;d(fg)=df g+(-1)^{s(f)}fdg \label{b15} \ee

These relations are consistent with the formula \be \partial_k \;\xi^i =\left[
ik \right]\;\xi^i\; \partial_k \label{b16} \ee

Basic operators $ i_k=\partial/\partial \xi^k $ of an inner derivation in the
algebra $ \Lambda_q (n) $ satisfy the following relations \bqa & i_k \;\xi^l
=\delta_k^l - \left[ lk \right]\; \xi^l\; i_k &\nonumber \\ & i_k \;x^l =
\left[ lk \right] \;x^l\; i_k & \label{b17} \\ & i_k\; i_l = -\left[ kl
\right]\; i_l\; i_k & \nonumber \eqa

A Lie derivative is the operator of zero degree in $ \Lambda_q (n) $ :  \be L_k
=L(\partial_k)=i_k\; d + d\;i_k \label{b18} \ee

The basic elements in a tangent vector space $ D_1 (M_q) $ are \be
x^{i_1}\;x^{i_2}\ldots x^{i_p}\;\partial_k =x^{I(p)}\;\partial_k =D_k^{I(p)}
\label{b19} \ee

Let us define a minimal deformation of commutator in $ D_1 (M_q) $ , which we
shall call the $ q $-commutator \be \left[ D_k^{I(p)},\; D_l^{J(r)}\right]_q
=D_k^I\; D_l^J - \left[ \frac {I|J}{k|l} \right]\; D_l^J \;D_k^I \label{b20}
\ee

It is not hard to verify that this $ q $-commutator can be decomposed in terms
of the sum of basic differential operators. Thus, the space $ D_1 (M_q) $ with
the operation (\ref{b20}) can be treated as minimal deformation of the
infinite-demensional Lie algebra $ Diff(R_n) $ . We shall use the name Lie $ q
$-algebras for the algebras of this type.

\setcounter{equation}{0} \section{Minimal deformation of the group GL(n) }
\indent

Consider a $ q $-commutative algebra $ A\; \mbox{over}\; {\bf C} $ with
generators $ a_k^i $ , which satisfy the relation \be \left[ a_k^i , a_l^j
\right]_q = a_k^i\; a_l^j -[ ij ]\;[ jk ]\; [ kl ]\;[ li ]\; a_l^j \;a_k^i =0
\label{c1} \ee

The algebra $ A(a^i_k) $ has a natural $ \widetilde{I}_q $-grading \be
\prod_{\alpha =1}^{p} \; a_{k_{\alpha}}^{i_{\alpha}} =a_{K(p)}^{I(p)}
 \rightarrow \;\; \left[ \frac{I(p)}{K(p)} \right]  \label{c2} \ee

Define a special tensor product $ A\bigotimes_q M_q $ as the $ q $-commutative
algebra with generators $ a_k^i \;\mbox{and}\; x^i $ \be a_k^i {\textstyle
\bigotimes_q} x^l - [ il ]\;[ lk ]\;x^l {\textstyle \bigotimes_q} a_k^i =0
\label{c3} \ee

Further we shall omit the symbol $ \bigotimes_q $ in multiplication of $ a_k^i
\;\mbox{and}\; x^l $.

Consider the $ I_q $-grading-conserving homomorphism $ \psi \;\mbox{of}\; q
$-commutative algebras which we shall call a linear cotransformation:  \bqa &
\psi :\; M_q (n)\; \rightarrow A {\textstyle \bigotimes_q} M_q (n) &\nonumber
\\ & \psi (x^i)=x^{\prime i}=a_k^i {\textstyle \bigotimes_q}\; x^k =a_k^i\; x^k
& \label{c4} \eqa

The widespread definition of linear cotransformation on $ M_q (n) $ is based on
using of the symmetrized tensor product

\bqa & \phi (x^i)=c_k^i \bigotimes x^k =x^k \bigotimes c_k^i & \nonumber \\ &
c_k^i\; c_l^j - [ ij ]\; [lk ]\; c_l^j\; c_k^i =0 & \label{c5} \eqa

This definition is not consistent with $ MD $-principle. Note, that a condition
on the components $ c_k^i $ can be written in the standard form $ R\; c_1\; c_2
=c_2\; c_1\; R $ , but a restriction on the components $ a_k^i $ in our
approach cannot be transformed to the standard form.

Denote a minimal deformation of the linear group by the symbol $ QGL(n) $.
Define a comultiplication $ \Delta $ in the algebra $ A(a_k^i )  $ \be
\Delta(a_j^i ) = a_k^i {\textstyle \bigotimes_q}\;a^k_j= [ ik ]\;[ kj ]\; [ ji
]\; a_j^k {\textstyle \bigotimes_q}\; a_k^i \label{c6} \ee

One can define also the following homomorphisms \bqa & s:\;\; A \;\rightarrow
\;A,\;\;\;\; s(a_k^i)\;a_l^k =\delta_l^i & \nonumber\\ & \varepsilon :\;\; A
\;\rightarrow \;{\bf C },\;\;\;\; \varepsilon (a_k^i) = \delta_k^i & \label{c7}
\eqa

It should be remarked that the homomorphisms $ \Delta , s \;\mbox{
and}\;\varepsilon \;$ conserve the $I_q $-grading. A standard system of axioms
in terms of commutative diagrams is valid for these homomorphisms. We shall
call the $q$-commutative bialgebras of this type by the name Hopf $q$-algebras.
Consider the $q$-commutative algebra of formal series $A(a^i_j , b^i_j )$. One
can introduce a formal group law , which corresponds to the comultiplication
 $\Delta $ \be \Delta (a^i_j )=\mu^i_j (a,b)= a^i_k \;b^k_j \label{c8} \ee

We shall study the automorphisms of the algebra $A(a^i_j )$ conserving
$\widetilde {I}_q$-grading. The simplest example of automorphism is connected
with the following substitution of generators:  \be a^i_j =\delta^i_j +
\alpha^i_j ,\;\;\;\;A(a)\;\rightarrow\;A(\alpha) \label{c9} \ee

This substitution change the parametrization of the formal group law and the
maps $ s,\;\varepsilon \;\mbox{in}\;QGL(n) $ \bqa & \mu ^i_j (\alpha ,\beta
)=\alpha ^i_j + \beta ^i_j + \alpha ^i_k \;\beta ^k_j & \nonumber \\ s^i_j
(\alpha )=\sum_{p=1}^{\infty}(-1)^p (\alpha ^p )^i_j  ,&\;\;& \;\varepsilon
(\alpha ^i_j )=0 \label{c10} \eqa

A validity of the $q$-group axioms can be verified in this parametrization \bqa
& \mu (\mu (\alpha ,\beta ),\gamma )=\mu (\alpha ,\mu (\beta ,\gamma ))&
\nonumber \\ & \mu (\alpha ,s(\alpha ))=\mu (s(\alpha ), \alpha)=0 &
\label{c11} \\ & \mu (\alpha , 0)=\alpha , \;\;\;\;\;\;\mu (0, \beta)=\beta &
\nonumber \eqa

In accordance with the results of Ref\cite{A8} there exists a correspondence
between deformations of formal Lie groups and deformed Lie algebras. It is
convenient to define a linear Lie $q$-algebra in terms of the fundamental
corepresentation of $ QGL(n) $ \bqa & \psi (x^i )=x^i + \alpha ^i_k\; x^k =x^i
+\delta (\alpha)\;x^i &\label{c12}\\ & \delta (\alpha)\;x^i=(\alpha^k_l \;L^l_k
)\;x^i &\label{c13} \eqa where $L^l_k $ is the linear operator in the vector
space spanned on generators $x^i$. The operator $\delta (\alpha)$ conserves the
$I_q$-grading, but the operator $L^l_k $ change this grading. In the simplest
representation $L^l_k $ has the form of the first-order differential operator
\bqa & L^l_k=x^l \;\partial_k &\label{c14} \\ & \left[L^i_k , L^j_m \right]_q
=\delta ^j_k\; L^i_m - [ij]\;[jk]\;[ki]\; \delta^i_m \;L^j_k &\label{c15} \eqa

The corresponding matrix representation has the following form \be L^i_k \;x^l
=(l^i_k )^l_j\; x^j =\delta ^l_k\; \delta ^i_j \;x^j \label{c16} \ee

A validity of the commutation relations (\ref{c15} ) for the matrices $l^i_k$
can be verified by a direct calculation. Denote by $qgl(n)$ the Lie $q$-algebra
with generators $L^i_k $ and the commutation relation (\ref{c15} ). Remark,
that the matrices $l^i_k$ can be treated as a representation of the Lie algebra
$gl(n)$ if the ordinary matrix commutator is considered instead of the
$q$-commutator.

    The contragradient representation of $qgl(n)$ can be realized on the
  algebra $M^*_q (y_i )$ \bqa & \widetilde{L}^i_k =-[kl] y_k \partial/\partial
  y_i &\nonumber\\ & \widetilde{L}^i_k \;y_l =(\widetilde{l}^i_k )_l^j \;y_j
  =-[kl]\;\delta^i_l\; y_k &\label{c17} \eqa

Define an action of the generators $L^i_k $ on tensor representations of
$qgl(n)$ by analogy with the action of the differential operator $ L^i_k
+\widetilde{L}^i_k $ on the monomials of the algebra $M_q (x)\bigotimes_q M^*_q
(y)$

\be L^i_k \;T^{i_1 \cdots i_p }_{j_1 \cdots j_r }\;\sim\;(L^i_k
+\widetilde{L}^i_k )\;x^{i_1} \cdots x^{i_p}\;y_{j_1}\cdots y_{j_r} \label{c18}
\ee

For example, a transformation of the tensor $T^{i_1 i_2}$ can be determined
with the help of a matrix \be (L^i_k )^{i_1 i_2}_{j_1 j_2}=\delta^{i_1}_k\;
\delta^i_{j_1}\;\delta^{i_2}_{j_2} + [ii_1 ]\;[i_1 k]\;\delta^{i_1}_{j_1}\;
\delta^i_{j_2} \;    \delta^{i_2}_k \label{c19} \ee

It is evident that an arrangement of indices has an essensial influence on
transformational properties of the $qgl(n)$-tensors.

It is easy to define the transformations of tensors by the operators of $\delta
(\alpha)$-type (\ref{c13}), which belong to a $q$-envelope of the algebra
$qgl(n)$. This object will be defined in section 4 by the analogy with a
Grassman envelope of Lie superalgebras [5].  \be \delta (\alpha)\;T^{i_1
i_2}=\alpha^{i_1}_i\; T^{ii_2} + \alpha^{i_2}_i\; T^{i_1 i} \;[ii_1]\;[i_1 i_2]
\label{c20} \ee

It should be stressed that the connection with the  $q$-algebra $qgl(n)$ arises
also in the standard approach to a definition of the coaction of the quantum
group $GL_q (n)$ on $M_q (n)$ (\ref{c15}) in terms of left-invariant
differential operators.

Note, that the algebra $qgl(2)$ is undeformed and isomorphic to the Lie algebra
$gl(2)$ in our approach. This fact was discovered in the investigation of a
two-parametric family of the $GL(2)$-deformations by a standard method [10].
It is easy to show that our version of differential calculus on $M_q (n)$
defined in section 2 is covariant under the action of $qgl(n)$ and $QGL(n)$.

A $q$-commutativity of the matrix elements $a^i_k $ (\ref{c1}) is the
sufficient condition for the preservation of $I_q$-grading under the linear
cotransformation (\ref{c4}) on $M_q (n)$. A necessity of the condition
(\ref{c1}) follows from the requirement of $QGL(n)$-covariance of a tensor
product $\bigotimes_q $ or relations (\ref{b3}),(\ref{b13}) and (\ref{b14}).

It is useful to consider an additional restriction on the deformation
parameters $q(i,k)=[ik]$, which leads to disappearence of deformations
$qgl(n)\rightarrow gl(n)$ for any $n$:  \be [ij]\;[jk]\;=\;[ik]\label{c21} \ee

The imposing of this condition retains $n-1$ independent parameters $q(i,i+1)$,
then the $q$-commutator (\ref{c15}) transforms into the ordinary commutator of
$gl(n)$. Even if we use the restriction (\ref{c21}) the action of
$gl(n)$-generator $L^i_k$ on $M_q$ depends on the parameters $[ik]$ and the
commutative elements $a^i_k$ of a formal $GL(n)$-transformation do not commute
with $x^l $.

\setcounter{equation}{0} \section{Minimal deformation of the matrix algebra}
\indent

In the theory of ${\bf Z_2}$-graded superspaces and supergroups there exists a
convenient for physical applications language of "points" which is equivalent
to the formulation in terms of bundles and Hopf superalgebras [6, 12, 13].
This approach establishes the correspondence of the linear supergroup $GL(p,
q)$ and the matrix algebra $\mbox{Mat}(p, q|\Lambda)$ with elements in some
arbitrary commutative superalgebra $\Lambda$ [5, 6].

We consider an analogous approach to a description of the $q$-deformed group
$QGL(n)$. Let $A^i_k (X)$ form an arbitrary set of elements depending on some
parameters $X$ and $Q(A)$ is the $q$-commutative algebra of formal series with
generators $A^i_k (X)$ \be \left[A^i_k (X)\; , \;A^l_m (X^\prime )\right]_q =0
\label{d1} \ee

Consider a set of $n\times n$ matrices $\mbox{Mat}(n, Q)$ with the elements in
$Q(A)$.  A structure of the matrix algebra $\mbox{Mat}(n, Q)$ can be defined
with the help of matrix addition and multiplication \bqa & (A+A^\prime )^i_k
=A^i_k (X) + A^i_k (X^\prime ) &\nonumber\\ & (A\;A^\prime )^i_k =A^i_l (X) \;
A^l_k (X^\prime ) &\label{d2} \eqa

One can multiply $A^i_k (X)$ by the central elements of the algebra $Q(A)$.
Note, that for the standard-type quantum matrices (\ref{c5}) $c^i_k
\mbox{and}\; c^{\prime i}_k $ addition and matrix multiplication cannot be
defined simultaneously.

There exists an uncertainty in the choice of $q$-matrix multiplication and one
can define the another operation \be (A\ast A^\prime )^i_k = A^l_k (X^\prime
)\; A^i_l (X) =[kl]\;[li]\;[ik]\; A^i_l (X)\;  A^l_k (X^\prime ) \label{d3} \ee

It is evident that these operations of matrix multiplication become identical
by the imposing of the condition (\ref{c21}) when the algebra $\mbox{Mat}(n,
Q)$ consists of matrices with commuting elements. In the general case we shall
use the definition (\ref{d2}).

The group $GL(n, Q)$ can be defined as a set of invertible matrices from
$\mbox{Mat}(n, Q)$. The elements $A^i_k (X) \in GL(n, Q)$ correspond to
homomorphisms of the Hopf $q$-algebra $QGL(n)$ to the algebra $Q(A)$ \be \phi_X
(a^i_k )=A^i_k (X) \label{d4} \ee A structure of the group $GL(n, Q)$
determines the homomorphisms $\Delta , s \;\mbox{and}\; \varepsilon $ in the
$q$-group $QGL(n)$.

Denote by $C(Q)$ a centre of the algebra $Q(A)$ and consider some maps
$\mbox{Mat}(n, Q) \rightarrow\;C(Q)$. The simplest map of this kind is
connected with a calculation of the trace of $q$-matrices \bqa & \Tr
A\;=\;A^i_i &\nonumber \\ & \left[ A^l_k (X)\;,\;\Tr A\right]=0 & \label{d5}
\eqa

The trace function on $q$-matrix products is invariant under the circular
permutation \be \Tr (A_1 A_2 \cdots A_p )=\Tr (A_p A_1 A_2 \cdots A_{p-1} )
\label{d6} \ee

One should use in the proof the ordinary commutativity of the elements $(A_p
)^k_i \; \mbox{and}\; B^i_k =(A_1  \cdots A_{p-1})^i_k $ with corresponding
values of indeces.

A standard form of the Hamilton-Cayley theorem [15] is valid for the matrix $
A\;\in\; \mbox{Mat}(n, Q)$ \be A^n =p_1 A^{n-1} + \cdots + p_{n-1} A + p_n
\label{d7} \ee where $p_r $ are coefficients of the characteristic polinomial
which can be written in terms of the functions $\Tr (A^k )$, for instance \be
p_1 =\Tr A \label{d8} \ee

A quantum determinant of the $q$-matrix A ( $q$-determinant ) by definition is
proportional to the coefficient $p_n $ \be \mbox{qdet}_n\; A=(-1)^n p_n
\label{d9} \ee

In the cases $ n=2,\;3$ quantum determinants have the following form:  \bqa &
\mbox{qdet}_2\; A\;=\;p_2 =\frac{1}{2}( \Tr A)^2 -\frac{1}{2}\Tr A^2
&\label{d10}\\ & \mbox{qdet}_3 \;A=-p_3 =\frac{1}{3} \Tr A^3 -\frac{1}{2}\Tr A
\Tr A^2 + \frac{1}{6}( \Tr A)^3 & \label{d11} \eqa

These formulas are completely equivalent to well-known expressions of $p_n $
for the complex matrices. In the same time the explicit formula of
$\mbox{qdet}_3$ in terms of the components $A^i_k $ contains an essential
distinction compared with a formula for the commutative $\mbox{det}_3$ :  \bqa
& \mbox{qdet}_3 \;A= A^1_1 A^2_2 A^3_3 - A^2_1 A^1_2 A^3_3 + A^2_1 A^3_2 A^1_3
-&\nonumber \\ & -A^3_1 A^2_2 A^1_3 - A^1_1 A^3_2 A^2_3 + A^1_2 A^3_1 A^2_3
\label{d12} \eqa

In comparison with det$_3 $ this formula has different order of elements in the
last term but the rest of terms are completely identical.

Now we present a formula for the function $\mbox{qdet}_n\;A $, which is useful
for a proof of the multiplicativity of this function in $\mbox{Mat}(n, Q)$. Let
us use for this purpose a linear cotransformation of the external form $dx^i
=\xi^i $ \be \xi^{\prime i}=A^i_k\; \xi^k \label{d13} \ee

Define a $q$-deformation of the antisymmetric symbol of $n$-th rank \be
(\varepsilon_q )_{i_1 i_2 \ldots i_n }=\prod_{1\leq \alpha < \beta }^{n}
\sqrt{\left[i_\alpha i_\beta\right]}\;\varepsilon_{i_1 i_2 \ldots i_n }
\label{d14} \ee where $\varepsilon_{i_1  \ldots i_n }$ is the ordinary
antisymmetric symbol.  Properties of $\varepsilon_q$-symbol follow from the
definition immediately, for example \be (\varepsilon_q )_{i_1 i_2 \ldots i_n }=
-[i_1\; i_2 ]\; (\varepsilon_q )_ {i_2 i_1 \ldots i_n } \label{d15} \ee

A contravariant symbol $(\varepsilon_q )^{i_1 \ldots i_n }$ can be defined by
analogy with (\ref{d14}). Consider the following identity:  \bqa &
(\varepsilon_q )^{j_1  \ldots j_n } \;(\varepsilon_q )_{i_n  \ldots i_1 }
\stackrel{\rm def}{=}\; n!\; \Pi^{j_1  \ldots j_n }_ {i_1  \ldots i_n } =&
\nonumber \\ & \delta^{j_1}_{i_1}\;\delta^{j_2}_{i_2}\cdots \delta^{j_n}_{i_n}
-[j_2 \;j_1]\;\delta^{j_2}_{i_1}\;\delta^{j_1}_{i_2}\cdots \delta^{j_n}_{i_n} +
 \cdots & \label{d16} \\ & \Tr \Pi =1 , \;\;\;\;\;\;\Pi^2 =\Pi & \label{d17}
\eqa where $\Pi $ is a deformation of the antisymmetric proectional operator of
the $n$-th rank .  \be \xi^{i_1}\cdots \xi^{i_n}= \Pi^{i_1  \ldots i_n }_ {k_1
\ldots k_n }\;\xi^{k_1} \cdots \xi^{k_n}=(\varepsilon_q )^{i_1 \ldots i_n }\;
V_n \label{d18} \ee

A constructive definition of the $q$-determinant is connected with the
transformation of the volume $n$-form $V_n$ in the group $GL(n, Q)$ \be
\xi^{\prime i_1 }\cdots \xi^{\prime i_n }=\mbox{qdet}_n\; A\; \xi^{ i_1 }\cdots
\xi^{ i_n } \label{d19} \ee

Using the Eqs (\ref{d13} , \ref{d18} ,\ref{d19}) and commutation rules of
$A^i_k \mbox{with}\; \xi^l $ we shall obtain the following relation \[
 (\varepsilon_q^\prime )^{i_1  \ldots i_n }= (\varepsilon_q )^{i_1  \ldots i_n
} \mbox{qdet}_n \;A = \] \be \prod_{\alpha =1}^{n-1}\; \prod_{\beta=\alpha
+1}^{n}[k_\alpha i_\beta ]\; A^{i_1}_{k_1}\cdots A^{i_n}_{k_n}\;(\varepsilon_q
)^{k_1  \ldots k_n } \label{d20} \ee

A formula for $\mbox{qdet}_n \;A $ can be obtained by contraction of this
relation with $(\varepsilon_q )_{i_n  \ldots i_1 }$. The Eq (\ref{d20}) is the
key relation for the proof of a multiplicativity of the function
$\;\mbox{qdet}_n  $ \be \mbox{qdet}_n (AB)=  \mbox{qdet}_n (A)\; \mbox{qdet}_n
(B) \label{d21} \ee

In this proof one use the following formulas \[ \prod_{\alpha =1}^{n}\;
 A^{i_\alpha }_{j_\alpha}\; B_{k_\alpha }^{j_\alpha} = \prod_{\alpha =1}^{n}
A^{i_\alpha }_{j_\alpha} \prod_{\beta =1}^{n} B^{j_\beta }_{k_\beta}
\prod_{\rho =1}^{n-1}\left[ \frac{j_\rho |i_{\rho +1}\ldots i_n }{k_\rho
		     |j_{\rho +1} \ldots j_n }\right], \] \be \left[
 \frac{j_\rho |i_{\rho +1}\ldots i_n }{k_\rho |j_{\rho +1} \ldots j_n }\right]
= \prod_{\lambda =\rho +1}^{n}\;\prod_{\sigma =1}^{n-1}\prod_{\gamma = \sigma
+1}^{n}\; [j_\rho i_\lambda ]\;[j_\lambda j_\rho ]\;[k_\sigma j_\gamma ]\;
[k_\gamma k_\sigma ] \label{d22} \ee

An exponential parametrization in the group $GL(n, Q)$ can be constructed on
the basis of Campbell-Hausdorf formula \bqa & e^{\textstyle u} \;e^{\textstyle
v} =e^{H({\textstyle u, v})} & \nonumber \\ & H(u,v)=u + v +\frac{1}{2}[u , v]
+\cdots & \label{d23} \eqa

This formula characterizes an exponential map of the Lie algebra $gl(n,
Q)=\mbox{Mat}(n, Q)$ to the group $GL(n, Q)$. The Lie algebra $gl(n, Q)$ is a
$q$-envelope of the the Lie $q$-algebra $qgl(n)$.

Other $q$-groups can be considered naturally as some subgroups of $GL(n, Q)$.
For instance, the group $SL(n, Q)$ can be determined by imposing of the
condition $\mbox{qdet}_n \;A=1 $. Denote $sl(n, Q)$ the corresponding Lie
algebra of traceless $q$-matrices $u$ \be \mbox{qdet}_n (e^{\textstyle u}
)=e^{\Tr \textstyle u} =1 \label{d24} \ee

The quantum group of formal diffeomorphisms Diff$(M_q (n))$ can be defined as a
set of formal homomorphisms of the algebra $M_q (n)$ preserving $I_q$-grading
\be x^{\prime i}=\sum_{p=0}^{\infty} \sum_{I(p)} a^i_{I(p)}(X) \;x^{I(P)}
\label{d25} \ee where $X$ is an arbitrary variety of parameters. It is evident
that this quantum group corresponds to the infinite-dimensional Lie $q$-algebra
$\;D_1 (M_q )$.

\setcounter{equation}{0} \section{Conclusion} \indent

Consider a partial example of the Zamolodchikov algebra \cite{A16} which is a
local analog of the algebra  $M_q (n)$:  \bqa & A^i (u)\; A^k (v)=q(i, k, u-v,
h)\; A^k (v)\; A^i (u)= & \label{e1} \\ & q(i, k, u-v, h) \;q(k, i, v-u, h)=1 &
\label{e2} \\ & q(i, i, 0, h) =1 ,\;\;\;\;q(i, k, u, 0) =1 & \label{e3} \eqa
where $q(i, k, u, h )$ are arbitrary functions, $u, v$ are the values of a
spectral parameter and $h$ is a quasi-classical parameter. An associativity of
the algebra (\ref{e1}) is equivalent to validity of the parametric quantum
Yang-Baxter equation (QYBE) for $\hat{R}(u) $ with arbitrary functions $q(i, k,
u, h) $.

It is shown in the quantum inverse scattering method (QISM) (see e.g.\cite{A14}
) that the quantum integrable systems correspond to the wide class of QYBE-
solutions $\hat{R}(u) $. It seems to us very interesting to use the solution
(\ref{e1}) in QISM.

  Note, that ${\bf Z_2}$-graded and $(G, f)$-graded generalizations of the
Yang-Baxter equation were discussed in Ref[17]. The physical applications of
the field commutation relations with constant $\hat{R} $-matrices  and some
examples of the diagonal $R$-matrices in QISM were considered in Ref[18].

The author would like to thank D.G.Pak for discussions.
 \setcounter{equation}{0}

 \end{document}